\documentclass[10pt]{iopart}

\usepackage{graphicx}
\usepackage{iopams}
\expandafter\let\csname equation*\endcsname\relax
\expandafter\let\csname endequation*\endcsname\relax
\usepackage{mathtools}

\usepackage[square,numbers,sort&compress]{natbib}

\DeclareMathOperator{\erf}{erf}

\begin{document}

\title[Burn regimes in the hydrodynamic scaling of perturbed inertial confinement fusion hotspots]{Burn regimes in the hydrodynamic scaling of perturbed inertial confinement fusion hotspots}
\date{\today}
\author{J K Tong, K McGlinchey, B D Appelbe, C A Walsh, A J Crilly, J P Chittenden}
\address{Centre for Inertial Fusion Studies, The Blackett Laboratory, Imperial College, London SW7~2AZ, United Kingdom}
\ead{j.tong15@imperial.ac.uk}

\begin{abstract}
We present simulations of ignition and burn based on the Highfoot and High-Density Carbon indirect drive designs of the National Ignition Facility for three regimes of alpha-heating \textemdash \ self-heating, robust ignition and propagating burn \textemdash \ exploring hotspot power balance, perturbations and hydrodynamic scaling.
A Monte-Carlo Particle-in-Cell charged particle transport package for the radiation-magnetohydrodynamics code Chimera was developed for this purpose, using a linked-list type data structure.

The hotspot power balance between alpha-heating, electron thermal conduction and radiation was investigated in 1D for the three burn regimes. The impact of perturbations on this power balance is explored in 3D using a single Rayleigh-Taylor spike.
Heat flow into the perturbation from thermal conduction and alpha-heating increases by factors of ${\sim}2-3$, due to sharper temperature gradients and increased proximity of the cold, dense material to the main fusion regions respectively.
The radiative contribution remains largely unaffected in magnitude.

Hydrodynamic scaling with capsule size and laser energy of different perturbation scenarios (a short-wavelength multi-mode and a low-mode radiation asymmetry) is explored in 3D, demonstrating the differing hydrodynamic evolution of the three alpha-heating regimes.
The multi-mode yield increases faster with scale factor due to more synchronous $PdV$ compression producing higher temperatures and densities, and therefore stronger bootstrapping of alpha-heating.
Effects on the hydrodynamic evolution are more prominent for stronger alpha-heating regimes, and include: reduced perturbation growth due to ablation from both fire-polishing and stronger thermal conduction; sharper temperature and density gradients at the boundary; and increased hotspot pressures which further compress the shell, increase hotspot size and induce faster re-expansion.
The faster expansion into regions of weak confinement is more prominent for stronger alpha-heating regimes, and can result in loss of confinement.

\end{abstract}
\noindent{\it Keywords}{: alpha-heating; ignition; burn; scaling; power balance; perturbed; inertial confinement fusion}
\\
\submitto{\NF}
\maketitle	

\ioptwocol
\section{Introduction}

Inertial confinement fusion (ICF) through indirect drive on the National Ignition Facility (NIF) involves laser illumination of a hohlraum to generate x-rays, which drive a spherical capsule implosion of deuterium-tritium (DT) fuel \cite{Lindl1995,atzeni2004physics}.
Inside the hotspot formed from the low-density gas in the centre of the capsule, alpha-particles are produced from DT fusion reactions.
The shell formed from the outer DT ice layer confines the hotspot while the alpha-particles deposit their energy through Coulomb collisions, heating the hotspot.
This heating induces more fusion reactions, producing the positive feedback process known as ignition which allows the hotspot to drive a propagating burn wave of fusion reactions through the rest of the DT shell, thereby resulting in high energy gain.

During the implosion, deviations from the spherical symmetry can be seeded by asymmetry in x-ray drive, inherent capsule surface roughness \cite{Smalyuk2015}, the tube used to insert the DT gas into the centre of the capsule \cite{Dittrich2016} and the support tent holding the capsule in the hohlraum \cite{Tommasini2015,Smalyuk2018}.
Perturbations can grow via instabilities such as Rayleigh-Taylor, allowing material from the outer layers of the capsule to penetrate and mix into the hotspot.
This can increase radiative losses from the hotspot due to high-Z material, and cool the hotspot due to mixing of hot and cold material.
Perturbations also reduce the efficiency of conversion of the shell's kinetic energy into thermal energy in the hotspot \cite{Scott2013}, increase the hotspot surface area and energy losses via thermal conduction \cite{Taylor2014}, truncate the burn pulse and can ultimately result in the loss of confinement \cite{Hurricane2016}.

Recent experimental progress on NIF has been encouraging, with the High-Density Carbon \cite{Divol2017,Pape2018} and Bigfoot \cite{Casey2018} campaigns delivering marked improvements from the Highfoot campaign in implosion symmetry and control \cite{Hurricane2014a,Clark2017}.
Previously, the radiation asymmetries and the tent-scar were the dominant degradation mechanisms \cite{Kritcher2016,Nagel2015}.
Their impacts have been reduced through a combination of improved laser drive symmetry control and different ablator materials \cite{Clark2018}.
Currently, the gas fill-tube is the most damaging perturbation source on the HDC implosions \cite{Pape2018}.
Methods for addressing the impact of the fill-tube \cite{MacPhee2018,Pape2018} have produced the highest yield implosions to date, with neutron yields of $\sim 2{\times}10^{16}$.

Hydrodynamic scaling of capsule size and laser energy has been explored as a method of investigating the potential driver requirements for indirect drive to achieve 1MJ of energy yield \cite{Clark2018a}.
Hydrodynamic scaling of a capsule design involves increasing the target dimensions and implosion time-scale by a factor of $S$, with other parameters in the setup remaining the same (e.g. the laser pulse is stretched by a factor $S$ in time, but remains identical in intensity).
This requires an increase in driver power and energy of $S^2$ and $S^3$ respectively. The hohlraum size also increases with $S$ in order to maintain an identical case-to-capsule ratio (CCR).
These hydrodynamically-equivalent implosions should have the same implosion velocities, adiabats and hydrodynamics \cite{Nora2014}, although non-hydrodynamic behaviour such as thermal conduction, radiation transport and alpha-heating do not scale in a similar way with implosion size \cite{Bose2015}.

In this paper, we provide a detailed study of alpha-heating under different potential perturbation scenarios, and how these scenarios change in different yield regimes.
Hydrodynamic scaling is used as a method of accessing the different yield regimes while keeping the implosions similar.
We first describe the progression through the three alpha-heating regimes of self-heating, robust ignition and propagating burn using scaled 1D simulations of the NIF Highfoot shot N130927 in section \ref{sec:1DIgnitionBurn}.
Next in section \ref{sec:3DPertn}, a 3D simulation of N130927 at $S=1.0$ of an idealised single-spike perturbation is used as an example to describe the impact of perturbations on the hotspot power balance and on the ignition process.
Section \ref{sec:3DScaling} then considers how more realistic perturbations might evolve as we progress towards increasingly more alpha-dominated by scaling up the HDC shot N161023 to larger capsule sizes, followed by conclusions in Section \ref{sec:Conclusions}.

All simulations in this paper are performed using the radiation-magnetohydrodynamics code Chimera \cite{Chittenden2016,Walsh2017,McGlinchey2018}, upgraded with the recently developed charged-particle transport module.
Chimera is a 3D Eulerian code which uses multi-group, $P_{\frac{1}{3}}$ automatic flux-limiting radiation transport \cite{McGlinchey2017,Jennings2005}; it uses a tabular equation of state calculated with the Frankfurt equation of state (FEoS) model \cite{More1988,Kemp1998,Faik2012}, and Epperlein-Haines \cite{Epperlein1986} modified Braginskii \cite{Braginskii1965} electron and ion thermal conductivities and equilibration rates.
Opacities and emissivities are interpolated using tables calculated using Spk \cite{Niasse2011}.
The charged-particle transport module uses a Monte-Carlo Particle-in-Cell model based on Sherlock's \cite{Sherlock2008} particle-fluid Coulomb collision model, with the Zimmerman \cite{Zimmerman1997} formulation of the Maynard-Deutsch \cite{Maynard1985} stopping model.
A detailed description of the module can be found in\ref{sec:AlphaModel}.

\begin{figure*}
\includegraphics{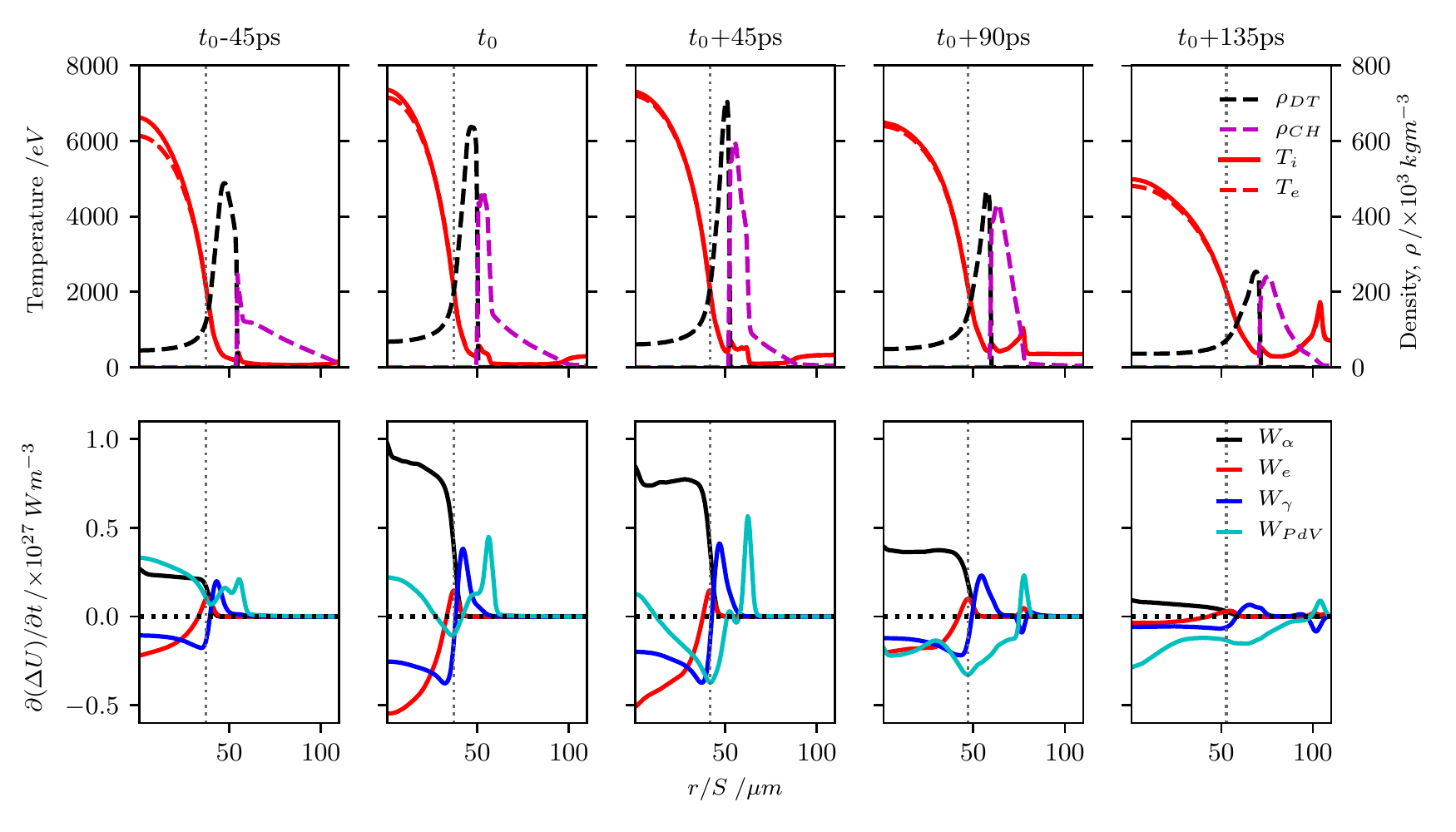}
	\caption{\label{fig:1D_NonIgniting} Time evolution of a 1D implosion in the self-heating regime based on shot N130927, at scale factor 0.9. Times are relative to peak compression at $t_0$. Densities and temperatures (above) are shown as fuel (dashed black) and ablator (dashed cyan) densities, ion temperature (solid red) and electron temperature (dashed red), while contributions to the hotspot power balance (below) are broken into alpha-heating, $W_{\alpha}$ (black), electron thermal conduction, $W_e$ (red), radiation, $W_{\gamma}$ (blue) and mechanical work, $W_{PdV}$ (cyan). The zero value is shown in the dashed black line, and the hotspot radius, $R_{hs}$ is given by the vertical dotted line. The alpha-heating is unable to increase hotspot temperatures after peak compression.}
\end{figure*}

\section{Ignition and Burn in 1D}\label{sec:1DIgnitionBurn}

\begin{figure*}
	\includegraphics{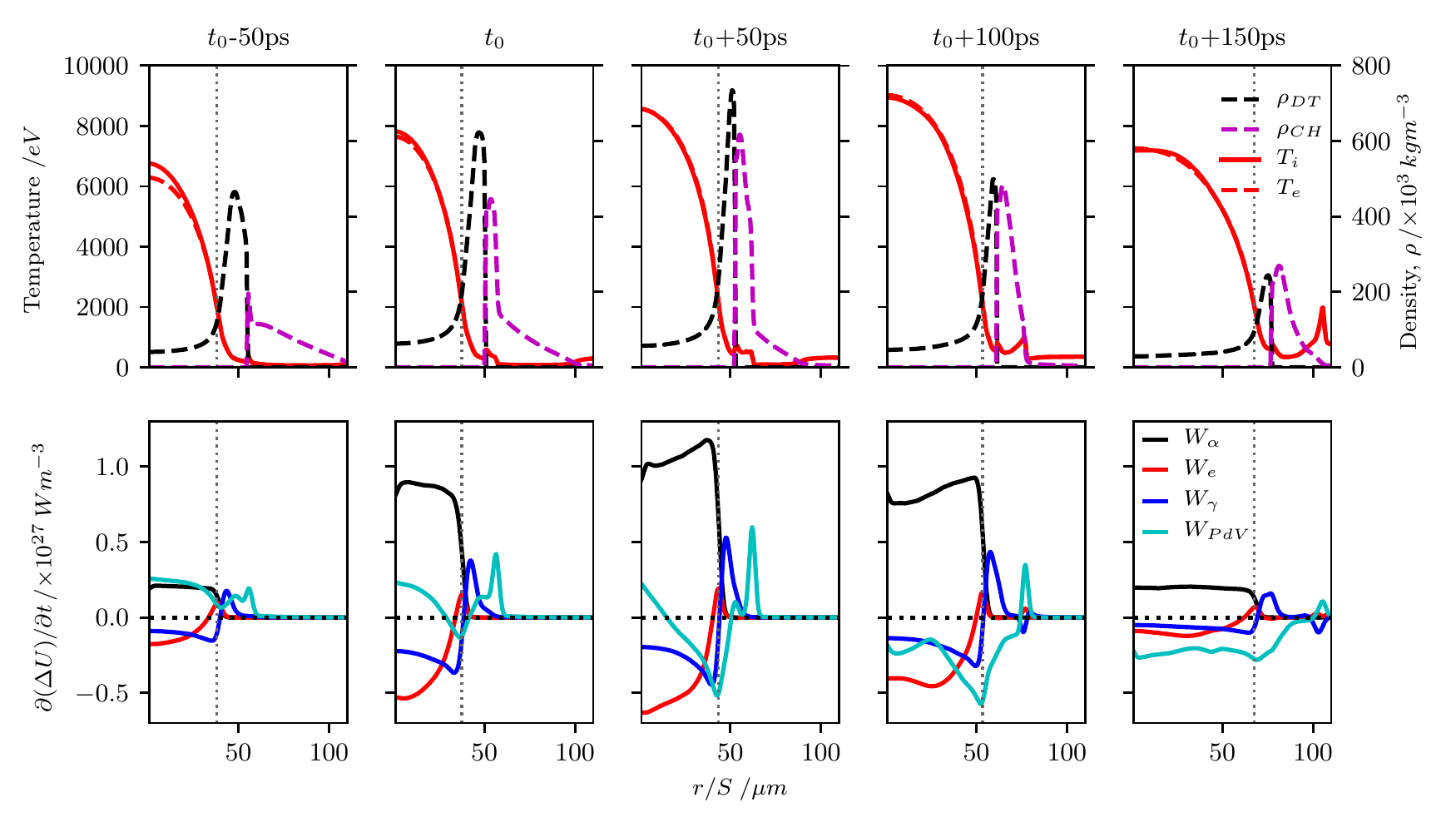}
	\caption{\label{fig:1D_RobustIgnition} Time evolution of a 1D implosion undergoing robust ignition, based on N130927 at scale factor 1.0 as shot. Quantities are as in figure \ref{fig:1D_NonIgniting}. Time-steps are proportional to scale factor to allow direct comparison between figures \ref{fig:1D_NonIgniting}, \ref{fig:1D_RobustIgnition} and \ref{fig:1D_PropagatingBurn}.}
\end{figure*}

\begin{figure*}
	\includegraphics{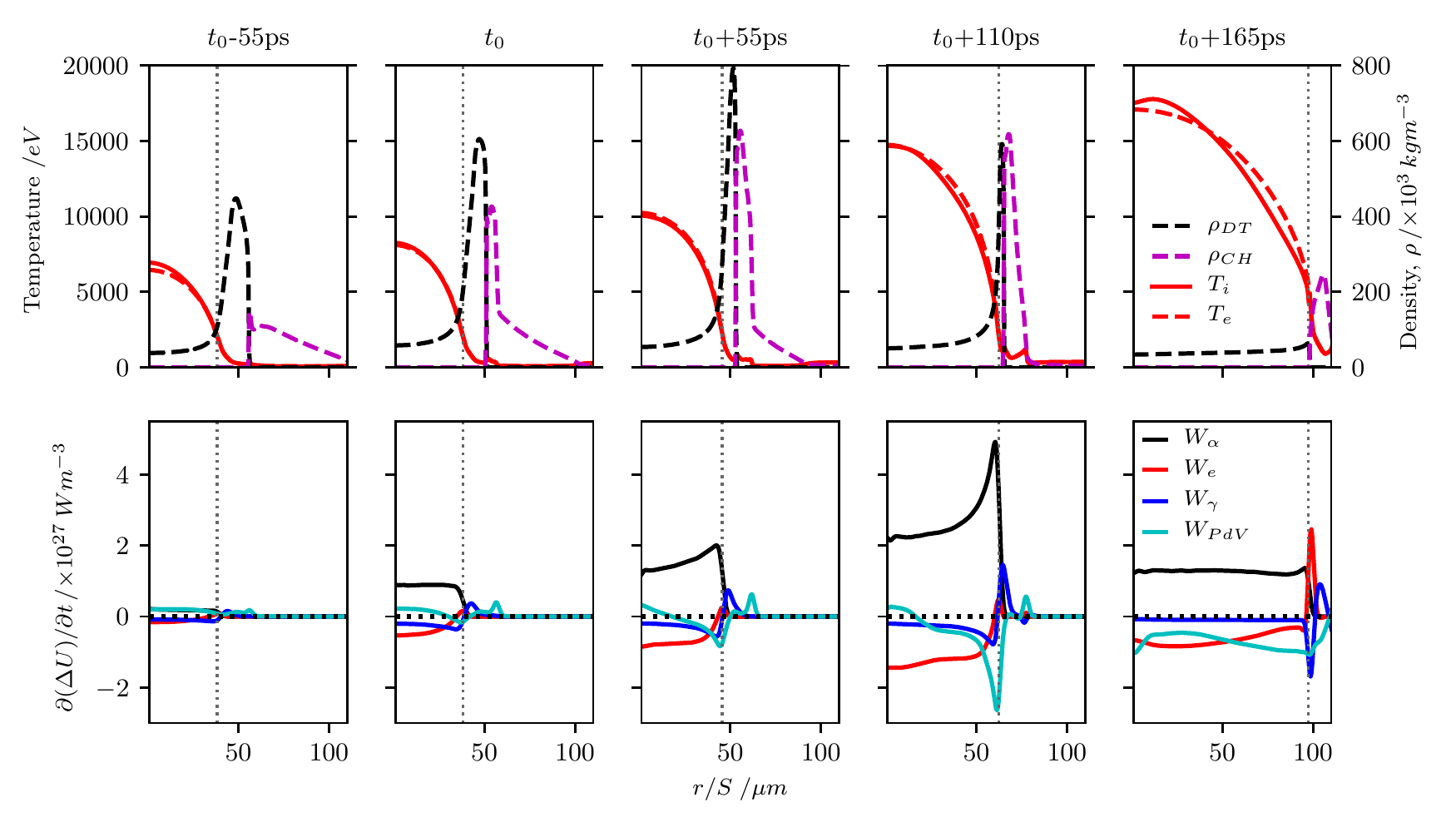}
	\caption{\label{fig:1D_PropagatingBurn} Time evolution of a 1D implosion in the propagating burn regime based on shot N130927, at scale factor 1.1. Quantities are as in figure \ref{fig:1D_NonIgniting}. Time-steps are proportional to scale factor to allow direct comparison between figures \ref{fig:1D_NonIgniting}, \ref{fig:1D_RobustIgnition} and \ref{fig:1D_PropagatingBurn}.}
\end{figure*}

In this section, we consider the impact of alpha-heating on the dynamics and evolution of the hotspot and illustrate the three regimes of ignition and burn which capsules can go through: self-heating, robust ignition and propagating burn.
To do this, we use scaled 1D simulations based on NIF Highfoot campaign shot N130927, one of the higher performing, well-studied Highfoot shots \cite{Hurricane2014}.
In the self-heating regime, low levels of alpha-heating deposit energy mainly in the hotspot, and the hotspot temperatures drop after peak compression as the alpha-heating is unable to compensate for the radiative, thermal conduction and expansion cooling.
We use the term `robust ignition' to describe the regime where stronger levels of alpha-heating boost the hotspot temperature significantly, with the resultant yield amplification due primarily to the increased fusion reactivity.
However, the confinement is not enough to allow a burn wave to develop.
Propagating burn results from the confinement of a robustly igniting hotspot, with the heat flow ablating significant proportions of the shell into the hotspot. The significant jump in fusion levels follows from the increase in hotspot density and therefore an increased fusion rate (proportional to $n_i^2$).

The simulations are run using Chimera in a spherical geometry at $500nm$ resolution.
The radiation transport algorithm uses 54 non-uniform radiation groups, chosen to provide sufficient resolution around features such as the carbon K-edge \cite{McGlinchey2018}.
These 1D simulations use a frequency-dependent x-ray drive spectrum, including the m-band component of the radiation flux \cite{McGlinchey2018}.

Figure \ref{fig:1D_NonIgniting} shows the time-evolution of a self-heating capsule, a scale 0.9 version of N130927.
Time is shown relative to peak compression (defined as the time $t_0$ when the spatially integrated $PdV$ work on the hotspot goes to 0), since this is the time after which alpha-heating is the only energy source within the hotspot.
The top panels display densities and temperatures, and the bottom panels the separate contributions to the hotspot power balance from alpha-heating ($W_{\alpha}$), electron thermal conduction ($W_{e}$), radiation ($W_{\gamma}$) and mechanical work ($W_{PdV}$).
Spatial scales are adjusted for scale factor, to allow easy comparison between regimes (see figures \ref{fig:1D_RobustIgnition} and \ref{fig:1D_PropagatingBurn} later).
The dotted line indicates the edge of the hotspot, as defined by the $2keV$ contour.

Here after peak compression ($t_0$), the contribution of alpha-heating towards the power balance is lower in magnitude than that of the combined losses from thermal conduction, radiation and $PdV$ expansion, and the hotspot temperature falls after peak compression due to the net power loss.
The alpha-particles spawning predominantly in the centre of the capsule travel a distance less than the size of the hotspot, such that the overwhelming majority ($\approx 85\%$) of the alpha-particle energy is deposited within the hotspot itself, as can be seen in the alpha-heating profiles throughout the burn pulse (bottom panels in figure \ref{fig:1D_NonIgniting}).
The transparency (the ratio of the average alpha-particle range, $l_{\alpha}$ to the hotspot radius $R_{hs}$) remains relatively constant in this regime over the period shown, with the alpha-particle energy deposited within the hotspot gradually increasing by $\sim5\%$ to $\sim85\%$.
As a result, in this self-heating regime the heating timescale is too long relative to the confinement time; the alpha-heating mostly heats the hotspot but not enough to compensate for the loss mechanisms, and contributes little to mass ablation of the shell.

Stronger alpha-heating levels result in the robust ignition regime, in which the alpha-heating has boosted the temperature of the hotspot significantly.
The yield increases significantly due to the increase in fusion reactivity, but there is not much ablation of fuel into the hotspot.
The hotspot mass increases only to around $35\%$ of the total fuel mass.
As a result, the burn-up of fuel is insignificant, with ${\Phi}  \lesssim 1\%$, where we define \cite{atzeni2004physics} the burn efficiency ${\Phi} = N_{fus} / N_{DT}^{(0)}$ for total number of fusion reactions, $N_{fus}$, and total number of DT pairs initially present, $N_{DT}^{(0)}$ (i.e. the total number of reactions possible).

Figure \ref{fig:1D_RobustIgnition} shows the time-evolution for a capsule in the robust ignition regime based on NIF shot N130927 at scale factor $S=1.0$, in the same format as figure \ref{fig:1D_NonIgniting}.
Timesteps between panels are equal when scale factor is adjusted for, to allow for direct comparison between figures \ref{fig:1D_NonIgniting}, \ref{fig:1D_RobustIgnition} and \ref{fig:1D_PropagatingBurn}.
Here, the `spark' from mechanical work on the hotspot allows for enough alpha-heating to increase temperatures slightly after peak compression, and maintain the temperature even against the capsule expansion.
However, the bootstrap is too weak to develop beyond heating just the hotspot; the areal density of the shell is unable to confine the hotspot in order to allow the heat flow to burn a significant proportion of the shell.
Here, the heating of the hotspot occurs on a similar timescale to the confinement of the hotspot.

We can observe the shift in the alpha-heating profile towards a Bragg peak.
This is due to the hotspot temperature increasing (from heating) and density decreasing (from expansion), and therefore becoming slightly more transparent (by about 10\% between $t_0$ and $t_0 + 100ps$), thereby allowing more alpha-particles to reach and thermalise in the dense fuel.
The coupling of alpha-particle energy into the hotspot drops from just above $80\%$ by $\approx5\%$ across the same time period.
The density-gradient scale length reduces by $\sim 20\%$, sharpening the density gradient at the hotspot-shell interface, due to a combination of the increased fuel ablation and the increased hotspot pressure.

For these first two regimes, as the level of alpha-heating increases, so too do the thermal conduction and radiative losses, both of which remain non-negligible throughout the burn pulse.
As expected, the hot centre of the hotspot drives the thermal conduction losses outwards, where it is reabsorbed in the inner region of the hotspot-shell boundary.
The radiative losses peak towards the outer edges of the hotspot where the density increases, but are reabsorbed in the dense shell (further out than where the thermal heat flow is absorbed).
It is worth noting that the hotspot is not optically thin, such that some of the bremsstrahlung is reabsorbed within itself (reducing the overall radiative power loss), and that a significant amount of the radiation is reabsorbed within the dense shell rather than escaping the capsule completely.
We could therefore consider some of this radiative energy as being recycled rather than purely lost (although indirectly, as to ablate and recycle the part of the shell that is heated by the radiation would require significant burn-up).

While robust ignition yield amplification is primarily due to increases in temperature, propagating burn yield amplification occurs as a result of ablating large proportions of the shell into the hotspot.
The significant boost in fusion levels follows from the contribution to the hotspot density (which would otherwise be falling due to hotspot re-expansion) and therefore to the fusion rate (proportional to $n_i^2$).
Maintaining the hotspot density at levels similar to peak compression while the hotspot is expanding and heating up results in high fusion rates being sustained for a much longer period of time.
As a result, propagating burn results in a significantly higher proportion of DT mass in the hotspot, and significant levels of burn-up, with ${\Phi} \sim 5\%$ ($Y_{DT}=1.7\times10^{18}$).

Figure \ref{fig:1D_PropagatingBurn} shows the time-evolution of the density, temperature and power balance of simulation of N130927 at scale factor $S=1.1$ undergoing propagating burn.
The temperature continues to increase even as the capsule expands, with the heating timescale faster than the confinement time.
The significant ablation and pressure increase which drive sharper gradients and faster hotspot re-expansion.

Only in this regime is the alpha-heating significantly greater than any of the loss mechanisms via thermal conduction, radiation or mechanical work.
The alpha-heating (black) in the bottom row shows the evolution from a sigmoid-like shape of the heating phase to the distinct Bragg-peak profile of a propagating burn, with the shell adequately confining the deflagrating hotspot against the substantial pressure build-up.
The Bragg peak evolves as a result of increasing alpha-range and stronger fusion production in close proximity to the shell, leading to more alpha-particles stopping abruptly in the dense fuel layer.
This strong deposition of energy leads to significant heating and ablation of material, allowing for substantial yield gains.

\begin{figure}
\includegraphics{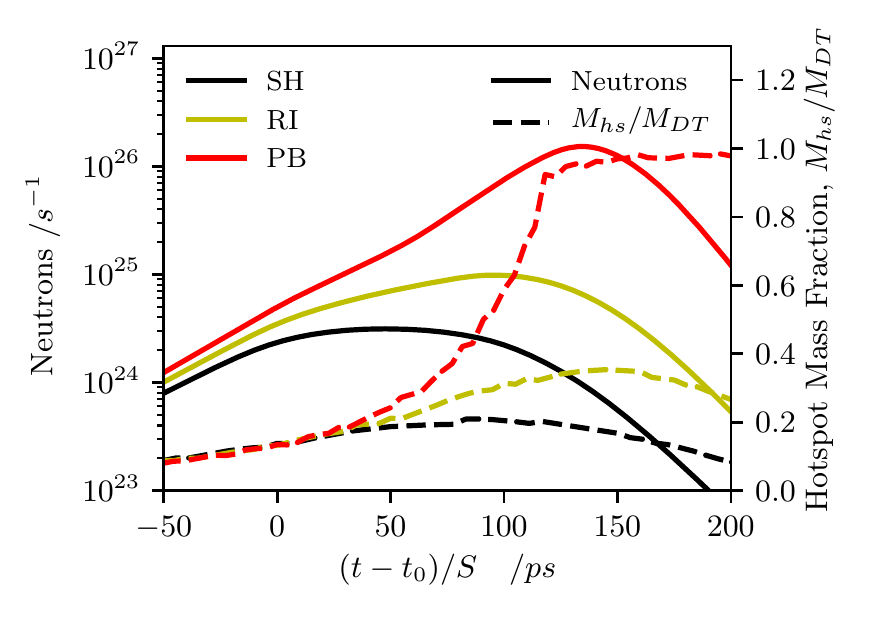}
	\caption{\label{fig:1D_RhoR_Mfrac} Neutron burn history (solid) and the fraction of DT mass in the hotspot ($M_{hs}/M_{DT}$) (dashed) against scaled time for the three regimes of alpha-heating shown in black (SH, self-heating), yellow (RI, robustly igniting) and red (PB, propagating burn).}
\end{figure}

Figure \ref{fig:1D_RhoR_Mfrac} shows the neutron burn histories and the fraction of DT mass contained within the $2keV$ hotspot ($M_{hs}/M_{DT}$) against time (adjusted for scale) for all three regimes.
The variation in hotspot mass with regime as a result of the differing levels of alpha-heating and heat flow can be seen clearly, with the self-heating hotspot containing up to only $20\%$ of the total DT mass, while the robustly igniting hotspot ablates more, up to around $30\%$.
The propagating burn hotspot expands to encompass almost the entirety of the DT fuel.

\section{Perturbations and heat flow}\label{sec:3DPertn}

While the hotspot ignites easily in an ideal, 1D scenario, the presence of inhomogeneities and asymmetries in 3D make ignition significantly harder.
In this section, we examine the impact of perturbations through their interactions with the heat flow from the hotspot, and vice versa, using the simplest case of a single dense perturbation spike applied to an otherwise spherically symmetric implosion of N130927 at $S=1.0$.
Methodologically, we simulate the drive phase (which constitutes the majority of the implosion) in 1D, and then use this data to initialise 3D simulations of the deceleration and burn phases at a convergence ratio of ${\sim}3$.
Here, we apply perturbations as Rayleigh-Taylor spikes in the velocity field to the capsule surface during the 1D-3D initialisation, following Layzer's \cite{Layzer1955} approximate analytic treatment of the instability \cite{Taylor2014, Chittenden2016}.
These 3D simulations do not include a radiation drive, and use a reduced 10-group structure in the radiation transport in order to expedite calculations while retaining accurate transport within the hotspot.

\begin{figure}
\includegraphics[scale=0.47]{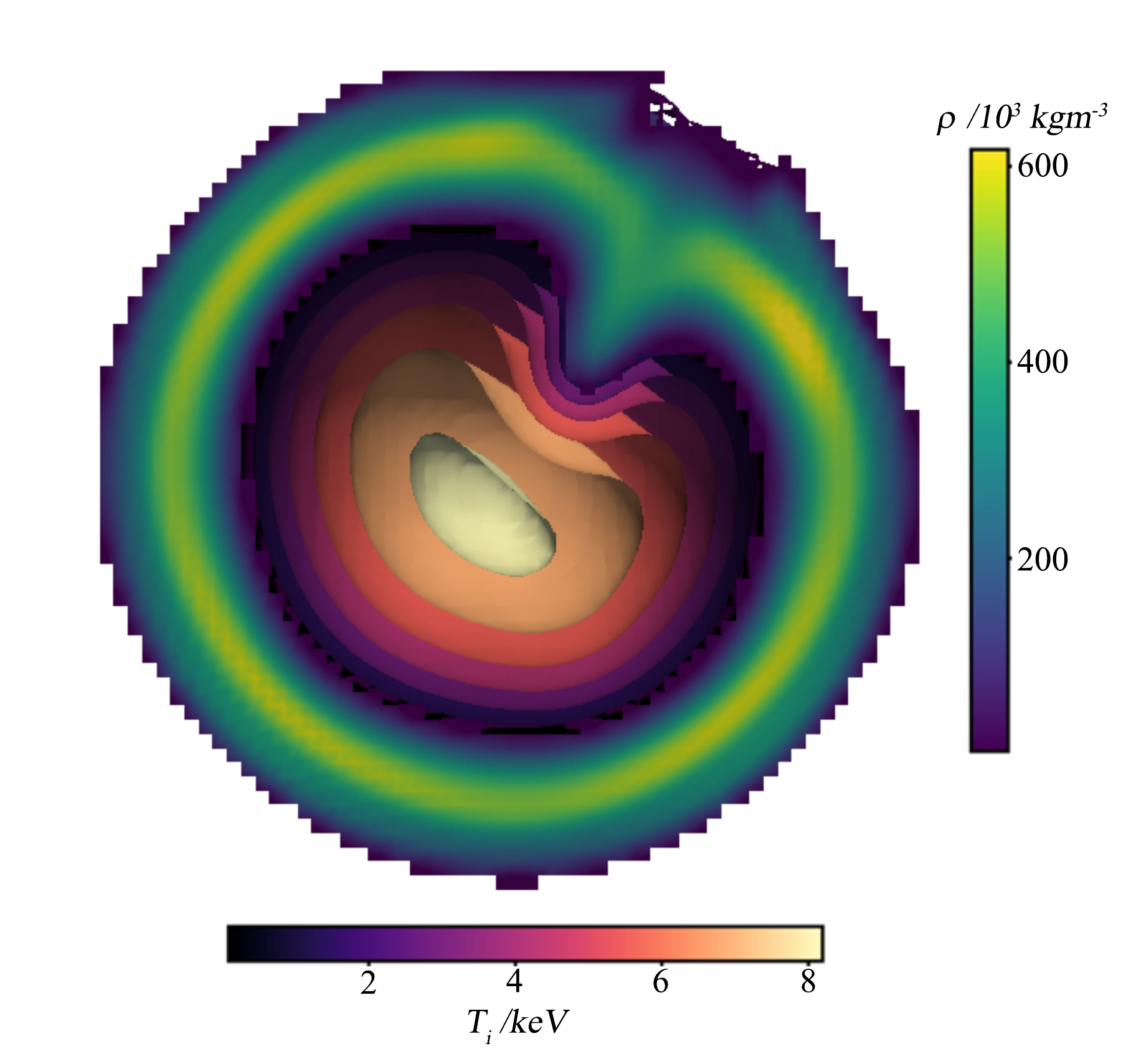}
	\caption{\label{fig:3DSS_RhoTi_1667} Ion temperature contours of the hotspot and a slice through the shell density at peak compression, for a single-spike perturbed implosion based on the Highfoot shot N130927 at scale factor $S=1.0$.}
\end{figure}

Figure \ref{fig:3DSS_RhoTi_1667} shows a 2D slice of the shell density, and ion temperature contours of the hotspot at peak compression. Even though the perturbation does not penetrate that deeply into the hotspot in density, the ion temperature distortion extends much deeper towards the centre of the capsule. 

\begin{figure}
	\includegraphics[]{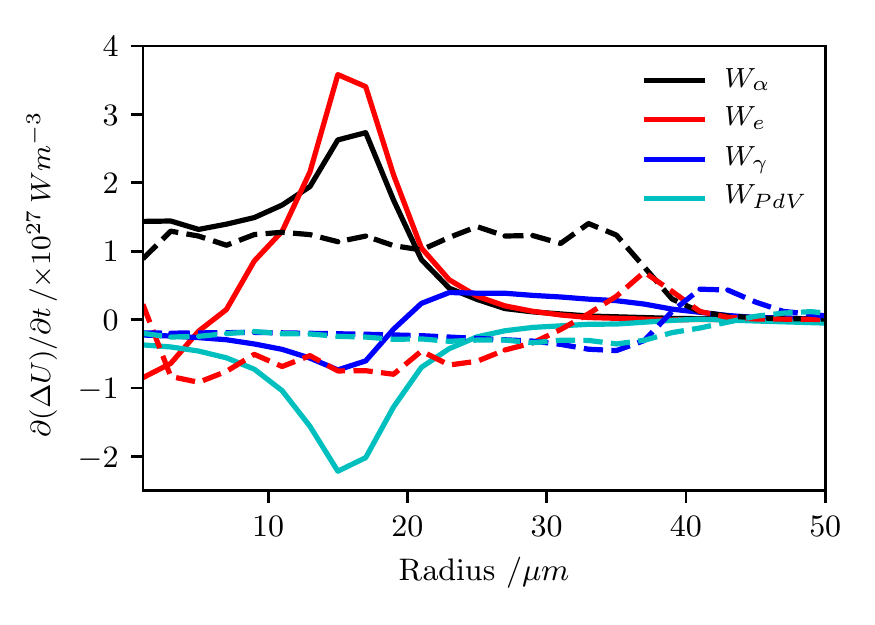}
	\caption{\label{fig:3DSS_PowersLoSs} Rates of change of energy density due to alpha-heating (black), thermal conduction (red), radiation (blue) and mechanical work (cyan) along radial lines of sight directly into (solid) and away from (dashed) the perturbation spike at peak compression.}
\end{figure}

Figure \ref{fig:3DSS_PowersLoSs} shows the impact of the spike on the hotspot power balance, comparing contributions from alpha-heating, thermal conduction and radiation along radial lines of sight directly into and away from the spike respectively.
The radiative contribution into the spike is similar to its counterpart away from the spike, apart from the reduced radial extent due to the spike penetrating deeper into the hotspot.
The alpha-heating into the spike peaks at roughly $3{\times}$ that of the unperturbed line of sight.
This increase in deposition is due to a significant flux of alpha-particles into the tip of the cold dense spike, whilst alpha-particles not encountering the perturbation deposit the majority of their energy as they transit across the hotspot, leaving little for the shell (as in the 1D scenario above for the robust ignition phase).
The sharpened temperature gradient (illustrated by compressed contours in figure \ref{fig:3DSS_RhoTi_1667}) around the spike also leads to increases of $<4{\times}$ in the thermal heat absorption compared to the rest of the shell.
The $PdV$ expansion work is much higher around the spike due to the increased heat flows causing more ablation.
Thermal conduction is the most significant contribution to heat flow into the spike, while alpha-heating is dominant for the unperturbed regions.

\begin{figure}
	\includegraphics[]{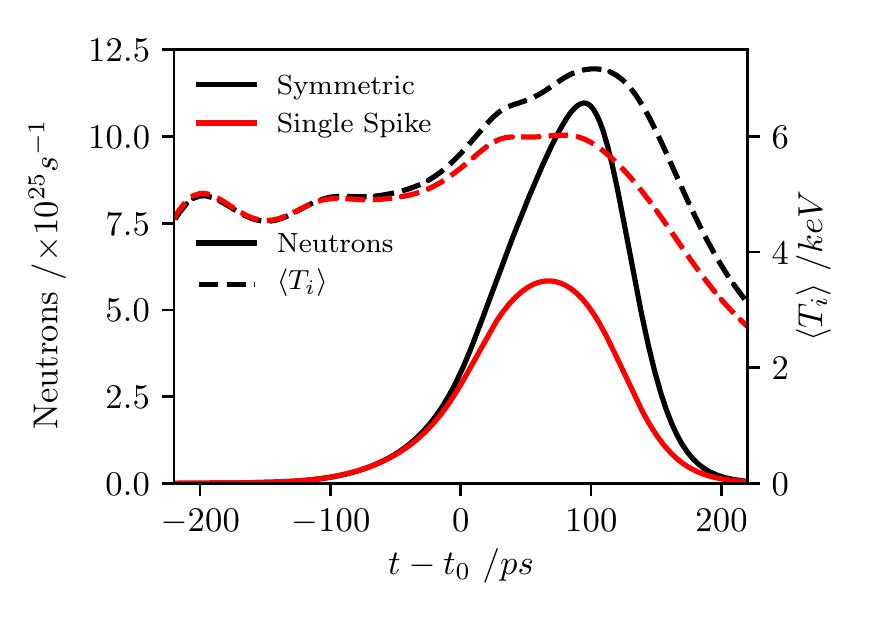}
	\caption{\label{fig:3DSS_P0_NeutronsBurnAvTi} A comparison of the burn pulse (solid) and burn-averaged ion temperature of the hotspot (dashed) between the unperturbed (black) and perturbed (red) implosions.}
\end{figure}

Figure \ref{fig:3DSS_P0_NeutronsBurnAvTi} shows the neutron burn history and the burn-weighted ion temperatures ${\langle}T_{i}{\rangle}$ for the perturbed single-spike implosion and an unperturbed, symmetric companion simulation.
The perturbation reduces the yield by roughly $40\%$, and results in a bang time $27ps$ earlier than for the symmetric case.
The transport of dense material towards the hotspot in this regime acts a cold sink for hotspot energy to reduce the ignition spark and quench burn, rather than as extra fuel to feed the burn.
Within the hotspot, ${\langle}T_{i}{\rangle}$ begins to diverge $50ps$ before peak compression, while the differences in the neutron pulses are visible only later on.
The differences in the early-time hotspot power balance manifest themselves through lower hotspot temperatures, particularly as the shell stagnates.
This is perhaps the most important time, as the mechanical work done during stagnation is converted into the hotspot thermal energy and therefore the spark from which the alpha-heating can bootstrap.
Since the synchronised stagnation of the shell for the symmetric case provides a higher rate of $PdV$ work and drives the hotspot temperature higher than for the perturbed case, the corresponding alpha-heating bootstrap is significantly stronger, and plays a large part in the differences in ${\langle}T_{i}{\rangle}$ and the fusion rate. 

Note that yield degradation from enhanced power losses from the hotspot and yield degradation from a weakened alpha-heating feedback loop due to early-time variations in hotspot conditions are difficult to distinguish

\section{Scaling perturbed simulations}\label{sec:3DScaling}

While implosion symmetry in recent years has improved remarkably, allowing yields to reach $2{\times}10^{16}$ \cite{Pape2018}, some level of perturbation persists.
With this view, we look to explore how the impact of different perturbation scenarios might scale to larger capsule sizes, and correspondingly higher laser energies.
As mentioned previously in Section \ref{sec:1DIgnitionBurn}, although the hydrodynamics of the capsule scales up identically, non-hydrodynamic phenomena such as thermal conduction, radiation transport and alpha-heating do not scale in a similar way, and are important factors regarding the growth and overall impact of the perturbations.
With perturbations in ICF implosions coming in various forms, do these different forms evolve and scale with capsule size differently from one another?

\begin{figure}
	\includegraphics[]{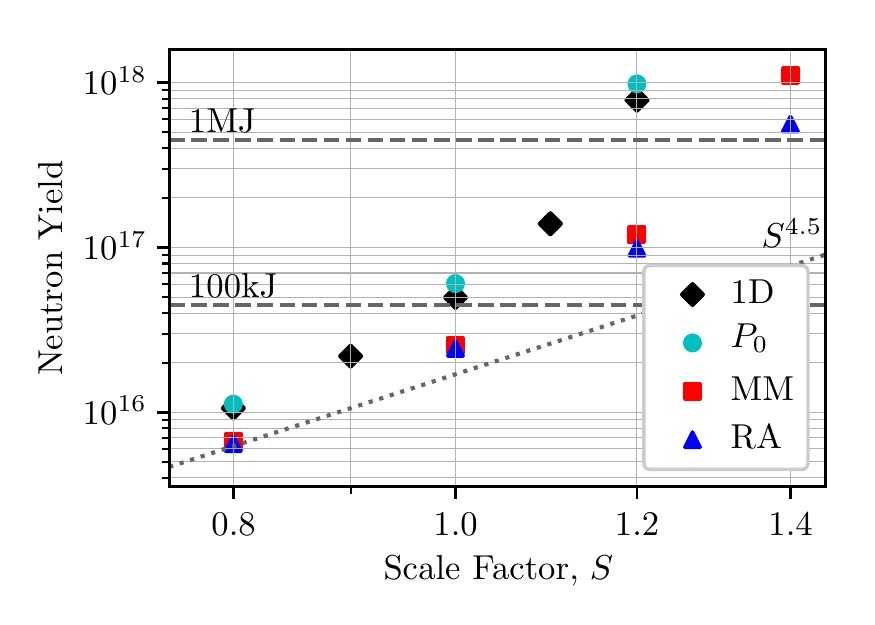}
	\caption{\label{fig:3DScaling_1D3DComp} Yields as a function of scale factor shown on a log-log scale for the high-resolution 1D (black diamond), the 3D symmetric $P_0$ (cyan circle), the multi-mode (red square) and the radiation asymmetry (blue triangle) scenarios. $1MJ$ and $100kJ$ yield levels are marked in dashes, and the 1D, no-alpha yield scaling ($S^{4.5}$) is shown as the dotted line.}
\end{figure}

\subsection{Methodology}

Simulations in this section are based on the NIF High Density Carbon (HDC) campaign shot N161023, a subscale 0.8 shot.
As in Section \ref{sec:3DPertn}, we simulate the ablation phase in spherical geometry, and then reinitialise onto a 3D Cartesian grid at peak radiation temperature.
The radial profiles of density, temperature and velocity are scaled in radius by the scale factor $S$ at this point of reinitialisation, such that all the simulations are based on the same initial dataset.

We describe two perturbation scenarios; multi-mode and radiation asymmetry.
The radiation asymmetry (RA) scenario is a low-mode shape asymmetry with Legendre modes $P_2$ and $P_4$ \cite{McGlinchey2018}, initialised from a 2D calculation with radiation drive asymmetries as derived from 2D hohlraum simulations \cite{Haan2017}.
The multi-mode (MM) scenario is initialised from a 1D calculation by applying Rayleigh-Taylor spikes in the velocity-field, as described in Section \ref{sec:3DPertn}.
Here, perturbations are applied at $N=42, 162, 642$ and $2562$ points uniformly distributed on a sphere, with wavelengths $R/\sqrt{N}$,  at a total of $3408$ points.
The MM perturbation amplitudes are tuned such that the $S=0.8$ yield matches that for the radiation asymmetry case.
The wavelengths of the MM perturbations scale with $S$, so as to remain the same size relative to the hotspot.

Due to the large number of simulations required for a scaling study, and the computational constraints imposed by resolution, we use simulations run at $3{\mu}m$ resolution.
We note that the lower burn-off yield of lower resolution 3D simulations also results in a reduced yield enhancement from alpha-heating, exacerbating the yield-loss due to resolution.
We account for this by applying a small ($4.8\%$) multiplier to the velocity at the 1D-3D reinitialisation, tuned such that the lower resolution symmetric 3D simulation yields match those of the high-resolution 1D simulations, verified at both $3{\mu}m$ and $2{\mu}m$.
Figure \ref{fig:3DScaling_1D3DComp} compares the yields for the 3D $3{\mu}m$ case (cyan circle) to the 1D case (black diamond), showing good agreement between the two.
The dotted line indicates the 1D, no-alpha yield scaling of $S^{4.5}$ \cite{Nora2014}.

\subsection{Performance Differences}

Figure \ref{fig:3DScaling_1D3DComp} also shows the yields for MM and RA.
Both scenarios display signs of ignition in the curvature of the yield graphs, even on the log-log scale.
All scenarios shown scale faster than the $S^{4.5}$ no-alpha, 1D hydrodynamic scaling.
Although the yields of MM and RA agree at $S=0.8$, MM scales better and faster, with a stronger ignition curvature.

\begin{figure}
	\includegraphics[]{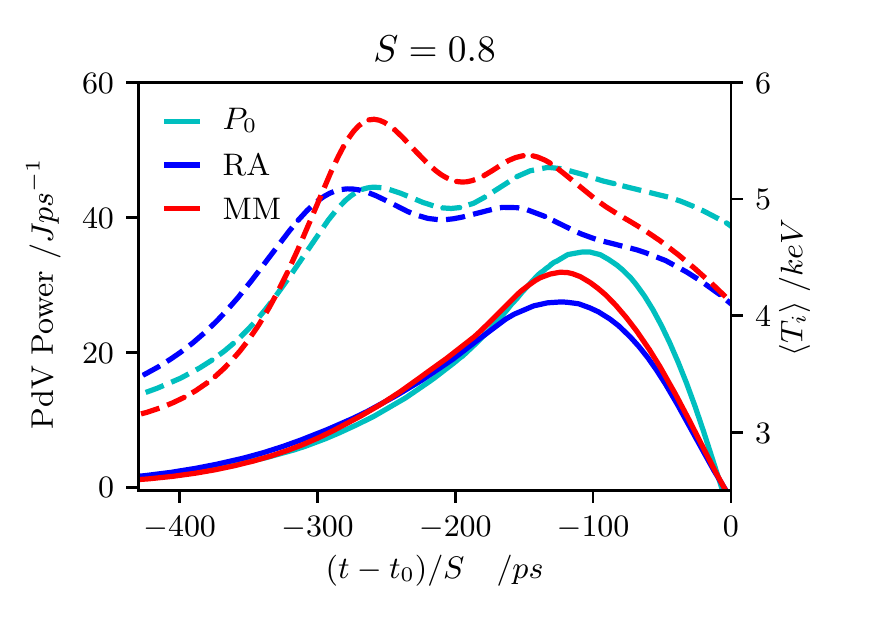}
	\caption{\label{fig:3DScaling_S08PdVTi} Hotspot burn-averaged ion temperature ${\langle}T_i{\rangle}$ (dashed) and $PdV$ power (solid) for $P_0$ (cyan), RA (blue) and MM (red) scenarios at scale factor $S=0.8$. The time relative to peak compression, $t-t_0$ is adjusted for the scale factor, $S$.}
\end{figure}

\begin{figure}
	\includegraphics[]{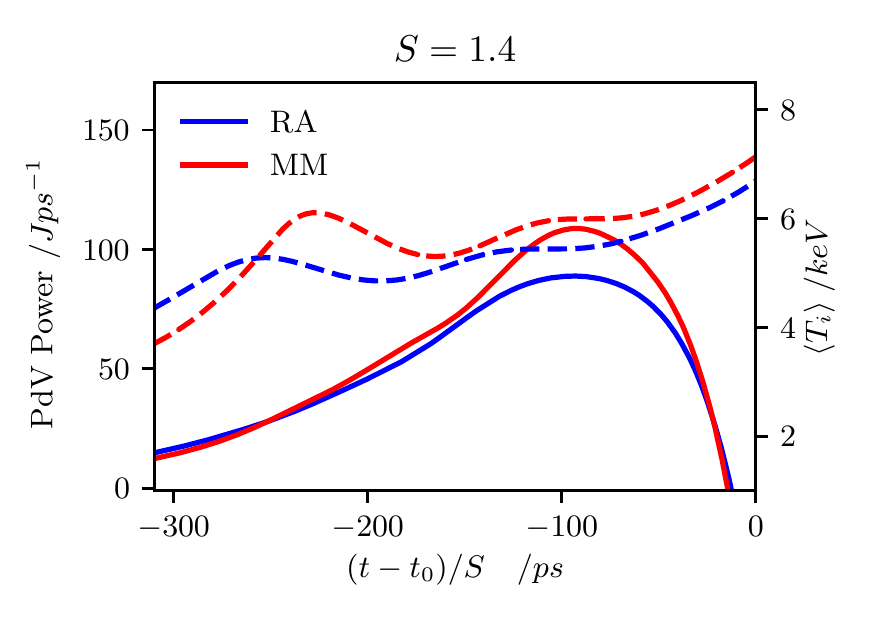}
	\caption{\label{fig:3DScaling_S14PdVTi} Hotspot burn-averaged ion temperature ${\langle}T_i{\rangle}$ (dashed) and $PdV$ power (solid) for RA (blue) and MM (red) scenarios at scale factor $S=1.4$.}
\end{figure}

\begin{figure}
	\includegraphics[]{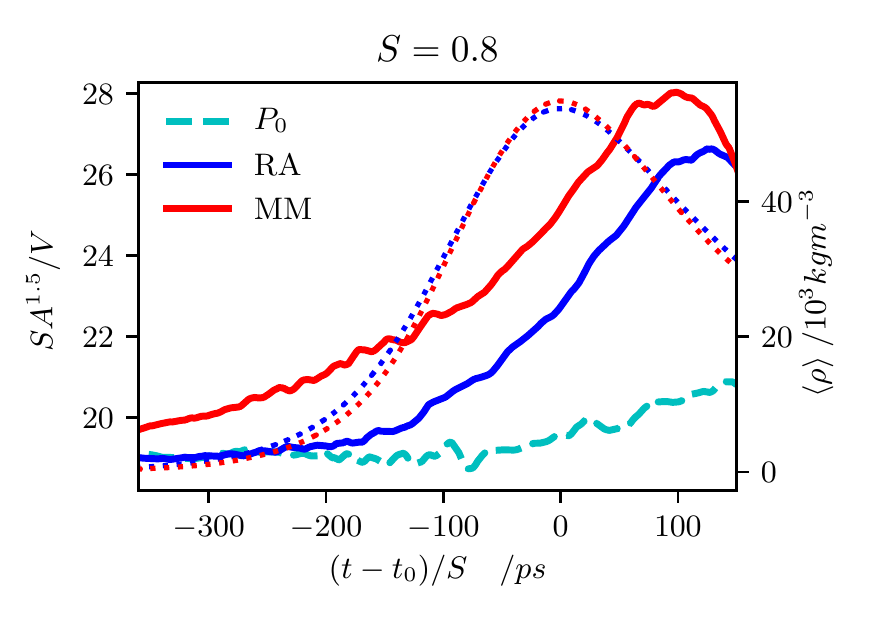}
	\caption{\label{fig:3DScaling_S08SAVolRatio}  Dimensionless measure of the surface area:volume ratio of the hotspot, $SA^{1.5}/V$ (solid) for RA (blue) and MM (red), with the symmetric $P_0$ case (dashed cyan) for reference. The hotspot burn-averaged density, ${\langle}{\rho}{\rangle}$ for both cases is also shown (dotted).}
\end{figure}

\begin{figure}
	\includegraphics[]{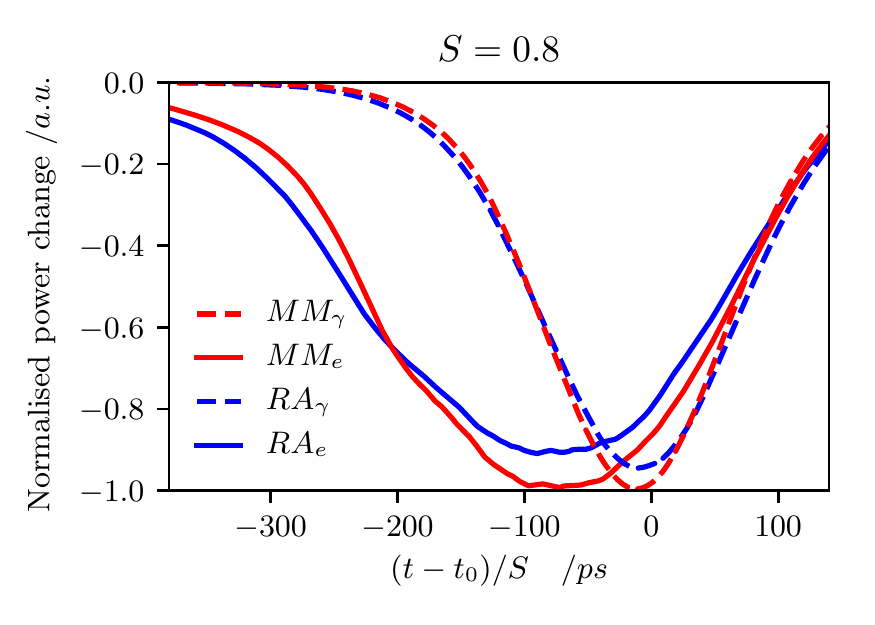}
	\caption{\label{fig:3DScaling_S08PowLoss} Normalised changes to the hotspot power balance for thermal conduction (solid) and radiation (dashed), for RA (blue) and MM (red) scenarios at scale factor $S=0.8$.}
\end{figure}

\begin{figure}
	\includegraphics[]{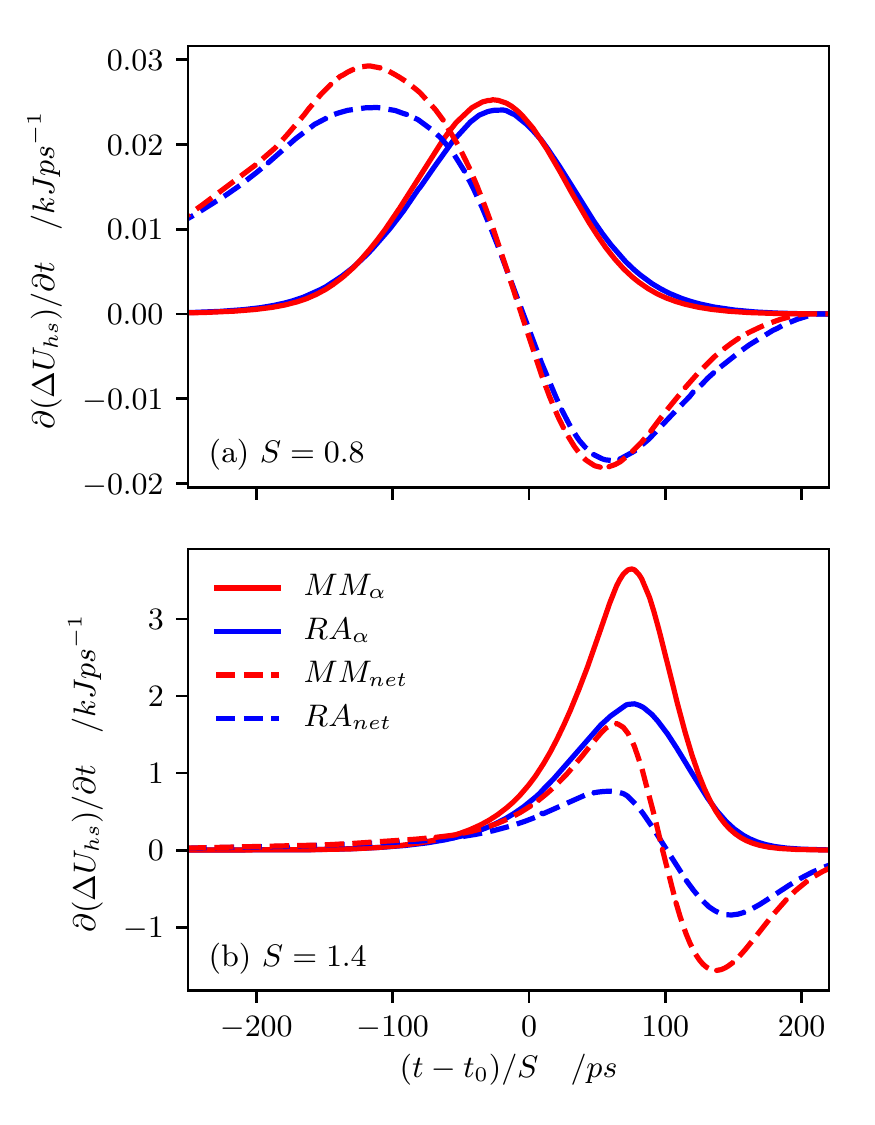}
	\caption{\label{fig:3DScaling_S0814PowBal} Hotspot power balance for (a) $S=0.8$ and (b) $S=1.4$, showing the net ($W_{net}$, dashed) and alpha-heating contribution ($W_{\alpha}$, solid) for RA (blue) and MM (red). $W_{net} = W_{\alpha} + W_{e} + W_{\gamma} + W_{PdV}$, where $W_i$ is positive for heating and negative for loss.}
\end{figure}

Figure \ref{fig:3DScaling_S08PdVTi} shows the $PdV$ power delivery to the hotspot for $P_0$, RA and MM, as well as the corresponding burn-averaged ion temperatures, ${\langle}T_i{\rangle}$, with times relative to the time of peak compression.
The MM has better $PdV$ delivery than RA, with the compressive work delivered in a more synchronous manner (based on the FWHM).
Physically, this is due to different parts of the shell stagnating at different times.
Due to the velocity differences, some regions of the shell in MM stagnate slightly earlier than other regions.
This effect is more prominent in RA, with the mass redistribution from the perturbation \cite{McGlinchey2018} causing larger momentum differences around the shell than for MM.
Higher $PdV$ power coincides with better heating of the hotspot.
The peak of the $P_0$ delivery is later than that of MM, but also higher and more synchronous, increasing the hotspot temperatures later in time but in a manner which maintains the temperature for longer.
When the $P_0$ delivery increases above that of MM, $d{\langle}T_i{\rangle}/dt$ also increases, while both quantities for RA remain below that of MM for the majority of the $PdV$ delivery.
This trend of more synchronous delivery of $PdV$ work follows to higher scales, resulting in higher peak-compression temperatures, as can be seen in figure \ref{fig:3DScaling_S14PdVTi}.

MM has stronger thermal conduction losses, driven in part by the higher temperatures and in part by a higher surface area (SA):volume (V) ratio.
Figure \ref{fig:3DScaling_S08SAVolRatio} shows the dimensionless ratio of $(SA)^{1.5}/V$ of the hotspot at $S=0.8$ for RA (blue) and MM (red), with that for the symmetric $P_0$ case shown for reference (cyan dashed).
The MM scenario has a consistently higher ratio, which contributes to the MM's increased thermal conduction losses relative to RA, which are shown in figure \ref{fig:3DScaling_S08PowLoss} alongside the marginal differences in radiative loss, in arbitrary units.
The higher hotspot density (dashed in figure \ref{fig:3DScaling_S08SAVolRatio}), combined with higher hotspot temperatures explains the higher radiative losses.
These higher losses cause similar performance between MM and RA at $S=0.8$ despite better $PdV$ compression.

At larger scales, alpha-heating becomes more significant in the hotspot power balance (and therefore the losses proportionally less so), which can be seen in figures \ref{fig:3DScaling_S0814PowBal}a and \ref{fig:3DScaling_S0814PowBal}b.
These figures compare the spatially-integrated alpha-heating contribution to the hotspot power balance, $W_{\alpha}$ (in solid) and the overall hotspot power balance, $W_{net} = W_{\alpha} + W_{e} + W_{\gamma} + W_{PdV}$ (in dashed).
This is due to the larger areal density of the hotspot and longer confinement time causing better absorption of alpha-heating energy.
As this occurs, the better $PdV$ compression of MM, which produces higher hotspot temperatures and densities, therefore produces stronger ignition than for RA.

\begin{figure*}
	\includegraphics[]{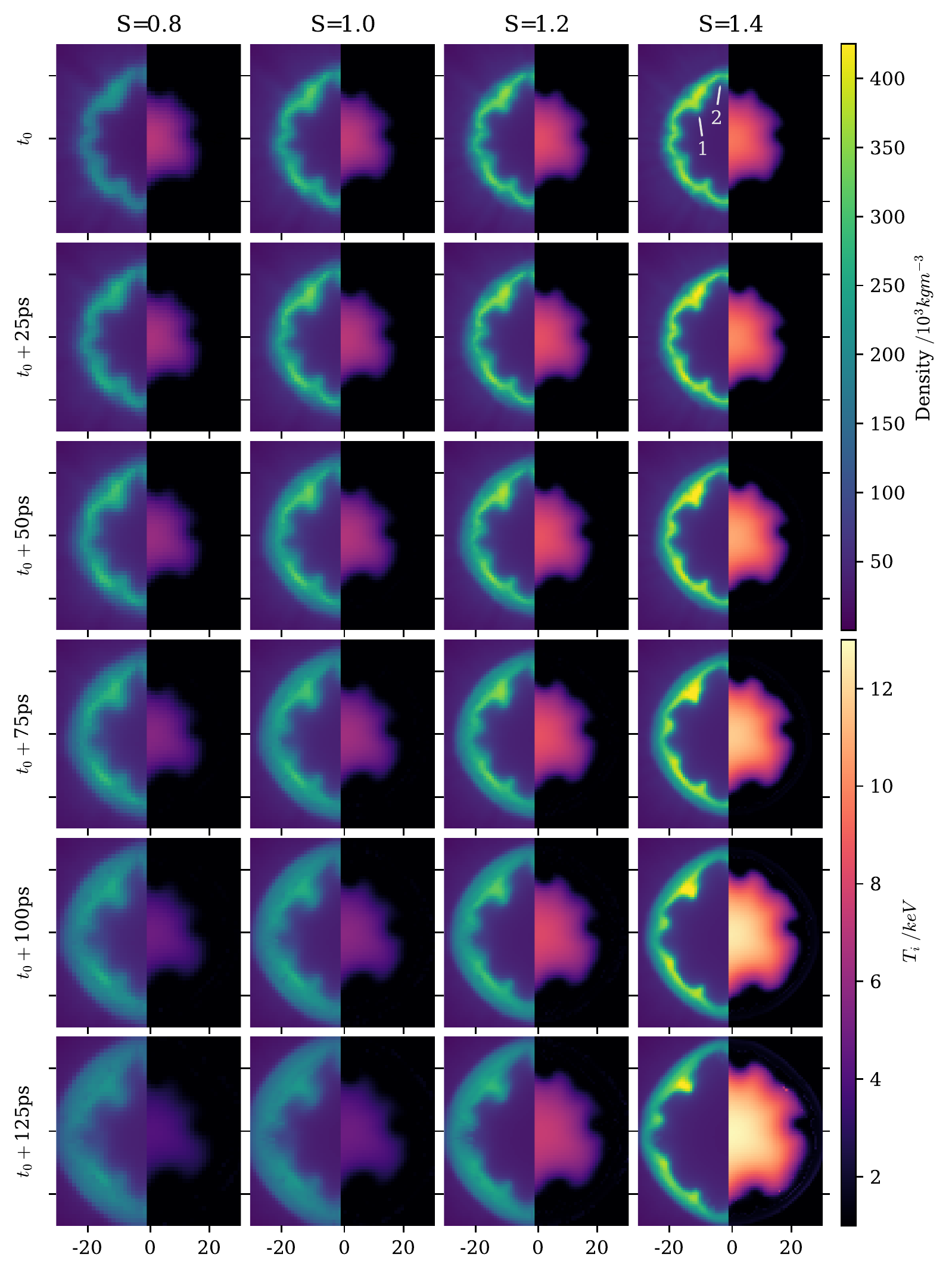}
	\caption{\label{fig:3DScaling_MM2DGrid} A grid showing the time-evolution (down the grid) of density (left half) and ion temperature (right half) slices for increasing scale factor, $S$ across the grid for the multi-mode scenario. The physical scale is normalised, $x/S \; ({\mu}m)$, in order to better compare the features across scale factors. Times are shown relative to the time of peak compression, $t_0$.
		 Annotations `1' and `2' indicate features exhibiting perturbation ablative stabilisation and bubble expansion.}
\end{figure*}
\begin{figure*}
	\includegraphics[]{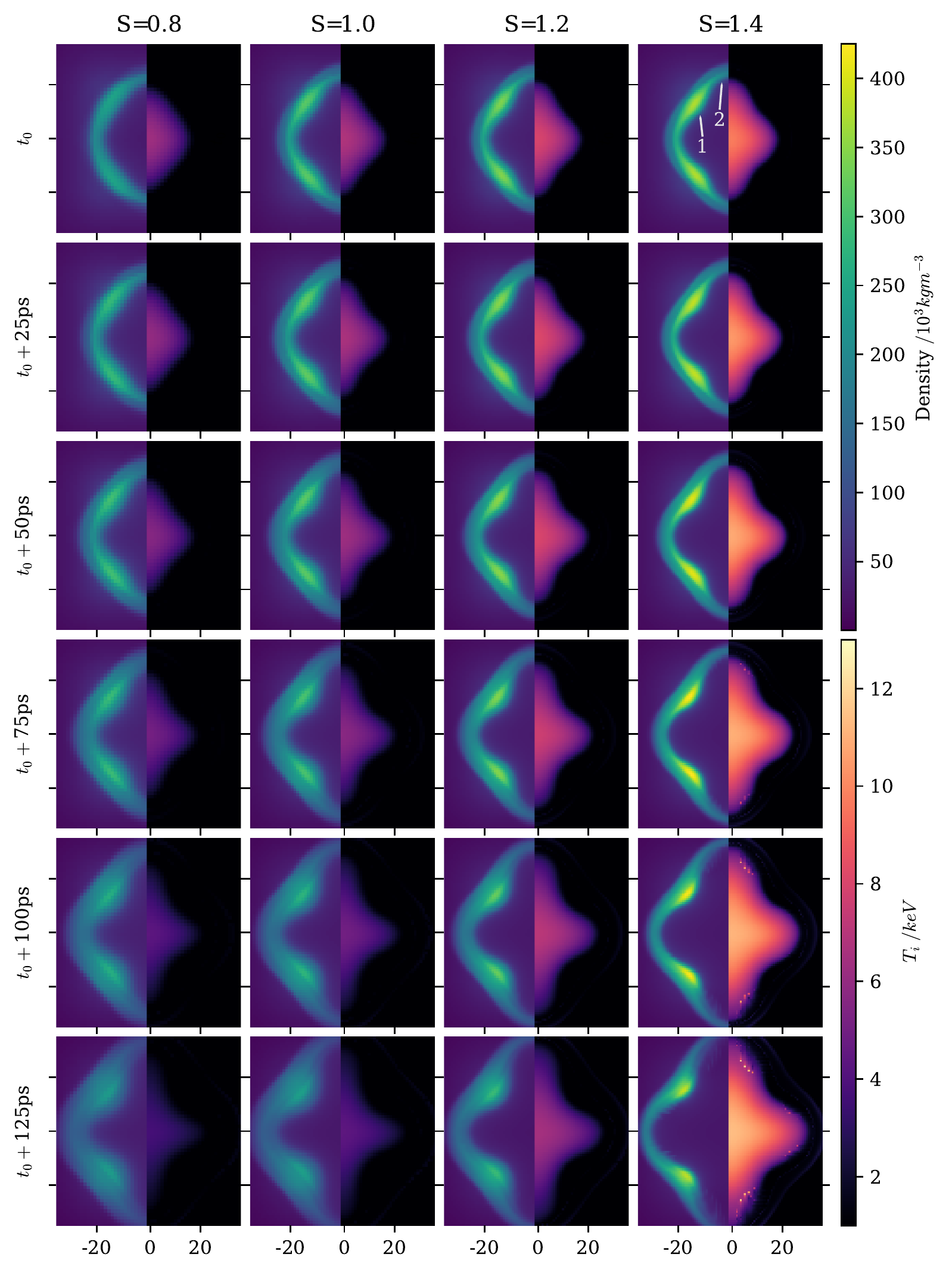}
	\caption{\label{fig:3DScaling_RA2DGrid} A time-evolution grid plot of density and ion temperature against scale factor as in figure \ref{fig:3DScaling_MM2DGrid}.}
\end{figure*}
\begin{figure*}
	\includegraphics[]{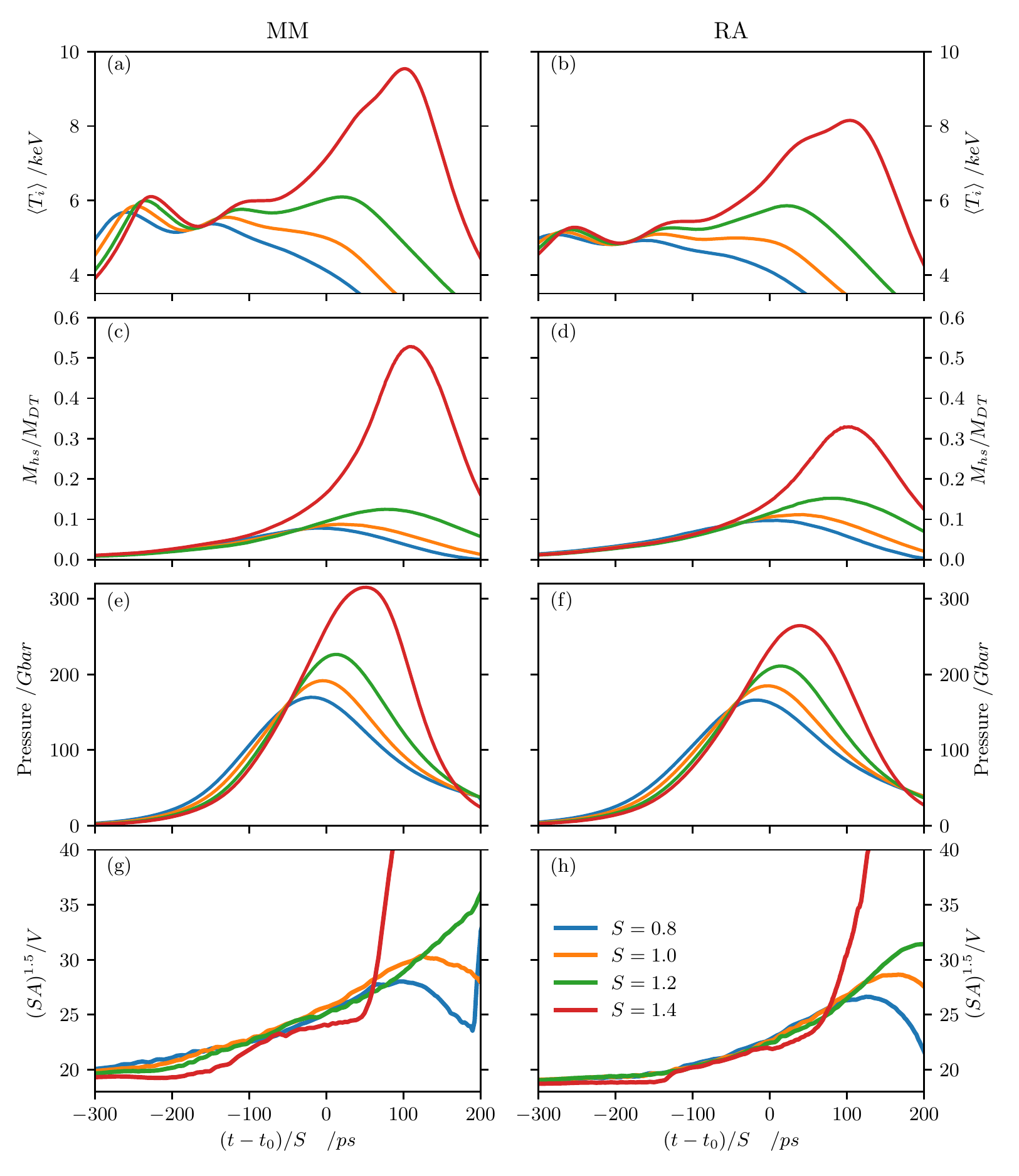}
	\caption{\label{fig:3DScaling_MMRATiMfracPressSAVr}(a,b) Burn-averaged ion temperatures, ${\langle}{T_i}{\rangle}$; (c,d) mass relative to total DT fuel mass, $M_{hs}/M_{DT}$; (e,f) hotspot pressure; and (g,h) dimensionless surface area to volume ratio, $(SA)^{1.5}/V$ of the hotspot for MM (left) and RA (right) respectively, across all four scales: $S=0.8$ (blue), $S=1.0$ (yellow), $S=1.2$ (red) and $S=1.4$ (green). Times are normalised by the time of peak compression, $t_0$ and adjusted for scale factor $S$.}
\end{figure*}

\subsection{Hydrodynamic effects of alpha-heating}

Figures \ref{fig:3DScaling_MM2DGrid} and \ref{fig:3DScaling_RA2DGrid} show the time evolution (down the grid) from peak compression (at $t=t_0$) of the MM and RA hotspots respectively as a function of scale factor (across the grid), displaying 2D slices of density (left half) and ion temperature (right half).
The physical scale is normalised by scale factor in order to better compare the hydrodynamics.
The three regimes from Section \ref{sec:1DIgnitionBurn} are demonstrated here using 3D results.
The side-by-side comparison of the time-evolution of these different regimes show clearly how the non-linear behaviour of alpha-heating causes differences in the hydrodynamic evolution, despite both scenarios being scaled up in a hydrodynamically self-similar manner.

Figure \ref{fig:3DScaling_MMRATiMfracPressSAVr} shows, for scenarios (MM,RA) respectively, the hotspot's: (a,b) burn-averaged ion temperatures, ${\langle}{T_i}{\rangle}$; (c,d) mass relative to total DT fuel mass, $M_{hs}/M_{DT}$; (e,f) pressure; and (g,h) dimensionless surface area to volume ratio, $(SA)^{1.5}/V$.

It can be seen that MM $S=1.4$ reaches the propagating burn regime, with figure \ref{fig:3DScaling_MMRATiMfracPressSAVr}a showing (in red) the continuing increase in hotspot temperature past peak compression, and figure \ref{fig:3DScaling_MMRATiMfracPressSAVr}c showing a significant proportion of the DT mass being ablated into the hotspot.
Meanwhile, RA $S=1.4$ shows a much weaker propagating burn, with a weaker (but still significant) temperature boost (red in figure \ref{fig:3DScaling_MMRATiMfracPressSAVr}b)and less DT mass ablation (figure \ref{fig:3DScaling_MMRATiMfracPressSAVr}d).
$S=1.2$ for both scenarios (in green) indicate a weak case of robust ignition, with temperature remaining relatively stable after capsule expansion and a relatively small change in $M_{hs}/M_{DT}$.
This indicates that the alpha-heating is strong enough to maintain the temperature against the $PdV$ expansion (but not enough to increase it), and relatively little mass ablation.
$S=0.8$ and $S=1.0$ for both scenarios (in blue and yellow respectively) fall within the self-heating regime, with temperatures dropping after peak compression and minimal change in the hotspot DT mass fraction.

Until around $100ps$ (scaled) before peak compression, figures \ref{fig:3DScaling_MMRATiMfracPressSAVr}a and \ref{fig:3DScaling_MMRATiMfracPressSAVr}b show the hotspot temperatures roughly agreeing with the hydro-scaled $\sim S^{0.2}$ \cite{Nora2014}.
However, from about $t_0 -50ps$, the hotspot ion temperatures begin to diverge from this scaling for the higher scales $S=1.2,1.4$.
This corresponds to the onset of significant levels of alpha-heating, with a stronger divergence for the stronger alpha-heating regimes shown in $S=1.4$. $S=1.0$ still agrees weakly, owing to relatively weak alpha-heating contributions.

While hydrodynamic scaling suggests constant pressure across scales \cite{Nora2014}, the raised hotspot temperatures also increase the hotspot pressures, as can be seen from figures \ref{fig:3DScaling_MMRATiMfracPressSAVr}a, \ref{fig:3DScaling_MMRATiMfracPressSAVr}b, \ref{fig:3DScaling_MMRATiMfracPressSAVr}e and \ref{fig:3DScaling_MMRATiMfracPressSAVr}f.
This in turn produces stronger back-compression of the confining shell, resulting in a thinner, higher density shell, and a hotspot larger than as scaled.
These effects are most visible in the right-most columns ($S=1.4$) of figures \ref{fig:3DScaling_MM2DGrid} and \ref{fig:3DScaling_RA2DGrid}.
Stronger heat flow into and ablation of the shell via thermal conduction and direct deposition of alpha-particle energy also contributes to the thinner shell, as well as to the higher temperature and density gradients at the hotspot-shell boundary visible here.

Perturbations are therefore better stabilised and growth reduced by this `fire-polishing' - an example of this can be seen at the location annotated `1' on figures \ref{fig:3DScaling_MM2DGrid} and \ref{fig:3DScaling_RA2DGrid}.
At the higher scales, we can see the perturbation here penetrates less deeply and recedes faster.
The effect of this is visible at earlier times in the plots of $(SA)^{1.5}/V$ in figures \ref{fig:3DScaling_MMRATiMfracPressSAVr}g and \ref{fig:3DScaling_MMRATiMfracPressSAVr}h.
Here, $(SA)^{1.5}/V$ is notably lower for $S=1.4$ (in red) in both cases, up until ${\sim}t_0 + 50ps$.
The fire-polishing effect reduces the perturbation growth and results in a more spherically-shaped hotspot than for the smaller scales (recall from figure \ref{fig:3DScaling_S08SAVolRatio} that the symmetric $P_0$ scenario has $(SA)^{1.5}/V \sim 20$).

Noting that the increased pressures will also result in faster hotspot expansion everywhere at higher scales, and that re-expansion into regions of weaker confinement (i.e. lower ${\rho}R$) will naturally occur faster than into regions of stronger confinement, we can see that the significantly increased pressures at $S=1.4$ amplify this difference in expansion rate.
The resultant `aneurysms' \cite{Hurricane2016} make the hotspot shape significantly less `1D'-like, and lead to the rapid rise in $(SA)^{1.5}/V$ after ${\sim}t_0+50ps$.
This coincides with the time of peak pressure, and the associated drop in pressure after this point.
This enhanced hotspot expansion can be seen in figures \ref{fig:3DScaling_MM2DGrid} and \ref{fig:3DScaling_RA2DGrid}.
Here, the locations annotated as `2' can be seen to expand faster than location `1' for higher scales.
Note that this expansion is shown to be faster in real time, not just scaled time, despite the fact that implosion timescales are also expected to increase by $S$.

The net effect of the faster re-expansion and increased heat flow on the hotspot energy balance can be seen from Figure \ref{fig:3DScaling_S0814PowBal}.
The power balance for MM at $S=1.4$ not only has a higher peak but also a lower trough, indicating not only stronger heating but also stronger losses.
The raised hotspot temperatures increase the thermal conduction and radiation losses, and the higher pressure increases the expansion losses.

\section{Conclusions}\label{sec:Conclusions}

We have developed a charged-particle transport model for Chimera which can be used to model the alpha-heating in inertial confinement fusion experiments.
We have used it to explore the hotspot power balance in ignition and burn, with three regimes of alpha-heating behaviour; self-heating, robust ignition and propagating burn.
In the self-heating regime, alpha-particles heat only the hotspot, with a long heating timescale relative to confinement, and no deflagration of shell material.
Alpha-particles boost the hotspot temperature significantly in the robust ignition regime, raising fusion yield due to a higher reactivity.
However, only in the propagating burn regime is there any significant ablation of shell material, requiring the hotspot to ignite robustly and be contained.
The fusion rate increases due to the increase (or rather, maintenance against expansion) of the hotspot density.

Hotspot heat flow losses due to thermal conduction and alpha-heating into an idealised, single-spike perturbation increased locally by ${\sim}2-3{\times}$ over the expected heat flow into the shell.
This is due to sharpened temperature gradients around the perturbation, and the increased proximity of fusion production regions to cold dense material resulting in a larger flux of alpha-particles into the perturbed shell. 
Part of the reduced performance compared to an unperturbed implosion is due to weaker bootstrapping from asymmetric shell stagnation.

We have explored the hydrodynamic scaling of a short-wavelength multi-mode and a low-mode radiation asymmetry scenario, and found that the multi-mode scenario yield scales faster and stronger than the low-mode scenario.
This is a result of more synchronous PdV compression work producing higher initial temperatures and densities, which at lower scales is compensated for by stronger hotspot losses and faster disassembly.
However, when scaled up to larger sizes, the higher areal density and longer confinement times allow for much more significant alpha-heating, which then bootstraps better for the higher temperatures and densities.
Hydrodynamic scaling appears to hold true for the first two regimes, but propagating burn exhibits the significant non-linearity in scaling behaviour due to alpha-heating.
Alpha-heating raises the hotspot temperature and pressure, resulting in stronger ablation and sharper gradients in density and temperature at the boundary.
Fire-polishing reduces perturbation growth, and the increased hotspot pressure back-compresses the shell to become thinner and higher density at peak compression.
The increased pressure results in larger hotspots and faster re-expansion, even more so than hydrodynamic scaling would suggest.
This increased pressure also causes significantly faster re-expansion into regions of weak confinement, resulting in loss of confinement through these regions at the highest scale factors.

We note that the inclusion of other, realistic sources of perturbation, such as the fill-tube might affect the ignition scaling.
In addition, other capsule and hohlraum designs currently being explored on NIF (such as Bigfoot \cite{Casey2018}, Beryllium \cite{Zylstra2018a}, HybridB) may be more robust to perturbations than the N161023-based design explored here.
Our investigation indicates that this specific design can indeed reach energy yields of 1MJ, even in the presence of large-amplitude perturbations.
However, the scale factors require are $\sim 1.3$ and $\sim 1.4$ for the multi-mode and radiation asymmetry perturbation scenarios respectively, corresponding ($\sim S^3$) roughly to laser energies of 4-5MJ.
		
\begin{ack}
The results reported in this paper were obtained using the UK National Supercomputing Service ARCHER and the Imperial College High Performance Computers Cx1 and Cx2.
This work was supported by the Engineering and Physical Sciences Research Council through grants EP/K028464/1, EP/M01102X/1, EP/L000237/1 and EP/P010288/1 as well as the Lawrence Livermore National Laboratory Academic Partnership Program and by AWE Aldermaston. 

\end{ack}

\appendix
\section{Alpha Model for Chimera}\label{sec:AlphaModel}

Accurate modelling of the alpha-heating within an ICF capsule requires non-local treatment, since the mean free path of $3.54MeV$ $\alpha$-particles is significantly larger than that of the background plasma. The alpha-particles undergo Coulomb collisions as they travel through the surrounding background plasma, and in doing so transfer energy to the electrons (and, to a lesser extent, the ions) \cite{Fraley1974}. Diffusive models \cite{Corman75, Honrubia1986} can handle the velocity-dependence of Coulomb collisions using energy-group structures and flux-limited diffusion methods to transport the particle energy. Diffusive models are not non-local, and are therefore less accurate in regions of high transparency; tending to overpredict the heating rate in the hotspot \cite{Taylor2013}, and struggling to recreate sharp Bragg peaks observed in regions of high stopping power. Particle-based models are more physical, but also more computationally-intensive; in 3D, large particle numbers are required for statistical and physical accuracy, as well as large quantities of random numbers and particle memory.

Modelling interactions between two charged particle species often uses a binary approach \cite{Takizuka1977, Nanbu1997}, in which particles within a cell are paired randomly and scattered off one another, conserving energy and momentum in the process. This is best suited for modelling multiple species as particles, since it models particle-particle interactions. However, a different approach is required for the particle-fluid interactions encountered in coupling the alpha model to the Maxwellian electron and ion fluids of Chimera. This approach, in which the collisional Fokker-Planck equation is reduced to Langevin form \cite{CADJAN1999, Jones1996a}, allows the use of fluid-based properties to calculate the slowing forces on particles.

\subsection{Numerical Implementation}

We follow the framework set out by Sherlock \cite{Sherlock2008} for implementing Coulomb collisions between particles incident on a Maxwellian background fluid moving at a finite velocity. The particle experiences both a deterministic frictional force and a stochastic velocity-space diffusion due to scattering. For a particle $\alpha$ of charge $Z_{\alpha}e$, mass $m_{\alpha}$ and velocity $v_{\alpha}$ scattering off background fluid $\beta$ of charge $Z_{\beta}e$, mass $m_{\beta}$, number density $n_{\beta}$, temperature $T_{\beta}$ and thermal velocity $v_{\beta}=\sqrt{2k_{B}T_{\beta}/m_{\beta}}$, the slowing (\ref{eq:SherlockSlow}) and diffusive (\ref{eq:SherlockDiffPara} and \ref{eq:SherlockDiffPerp}) coefficients are given \cite{Sherlock2008}:
\begin{subequations}
	\begin{align}
	\frac{\partial v_{\alpha\parallel}}{\partial t} &= -A(1+m_{\alpha}/m_{\beta})\frac{G(x)}{v_{\beta}^{2}}\label{eq:SherlockSlow} \\
	\frac{\partial v_{\alpha\parallel}^{2}}{\partial t} &= AG(x)/v_{\alpha}\label{eq:SherlockDiffPara} \\
	\frac{\partial v_{\alpha\perp}^{2}}{\partial t} &= A\frac{\erf(x)-G(x)}{v_{\alpha}} \label{eq:SherlockDiffPerp}
	\end{align}
\end{subequations}
where  $x=v_{\alpha}/v_{\beta}$, ${\erf(x)}$ is the error function, $\ln\Lambda_{\alpha\beta}$ is the Coulomb logarithm between a test particle $\alpha$ scattering off field particles $\beta$, and $A$ and the Chandrasekhar function $G(x)$ are given by:
\begin{align}
A &= \frac{Z_{\alpha}^{2}Z_{\beta}^{2}e^{4}n_{\beta}\ln\Lambda}{2\pi m_{\alpha}^{2}\epsilon_{0}^{2}} \\
G(x) &=\frac{\erf(x)-x\frac{\partial\erf(x)}{\partial x}}{2x^{2}}
\end{align}
The framework allows for exact conservation of energy and momentum. Conservation of momentum from individual particles crossing the grid cell can be used to calculate the change in fluid momentum and the change in fluid kinetic energy ($\Delta K_{fluid}$), and conservation of energy used to calculate the change in total fluid energy ($\Delta E_{fluid}$). This then gives the change in thermal energy in the fluid, $\Delta U_{fluid}=\Delta E_{fluid}-\Delta K_{fluid}$. The calculations are expedited by using the following approximation for $G(x)$ \cite{Sherlock2008}: 
\begin{equation}
G(v_{\alpha}/v_{\beta}) \approx \frac{v_{\alpha}v_{\beta}^{2}}{2v_{\alpha}^{3}+\frac{3\sqrt{\pi}}{2}v_{\beta}^{3}}
\end{equation}
Particle motion is integrated using a leapfrog scheme, accurate to second order in time \cite{Birdsall1985}, and subcycled according to the particle cell-crossing time and the slowing (${\tau}_s$) and scattering (${\tau}_{\parallel}, {\tau}_{\perp}$) relaxation times:
\begin{equation}
\tau_{s}=\frac{v_{\alpha}}{\partial v_{\alpha}/\partial t},\:\tau_{\parallel}=\frac{v_{\alpha}^{2}}{\partial v_{\alpha\parallel}^{2}/\partial t},\:\tau_{\perp}=\frac{v_{\alpha}^{2}}{\partial v_{\alpha\perp}^{2}/\partial t}
\end{equation}
Particles are spawned at random locations within each cell with a broadened energy distribution \cite{Appelbe2011}, and an isotropic velocity distribution.
Fuel energy \cite{Clayton1983} and mass are removed from the cell based on the number of fusion reactions in the cell.
While alpha-alpha collisions are not included in this scheme, these collisions are not expected to be significant as the alpha-particle number density is low.
We also do not include a helium ash fraction for the thermalised alpha-particles, since burn-up is generally very low (typically $\lesssim 1\%$, at most $\lesssim 5\%$).

\begin{figure*}
	\includegraphics[]{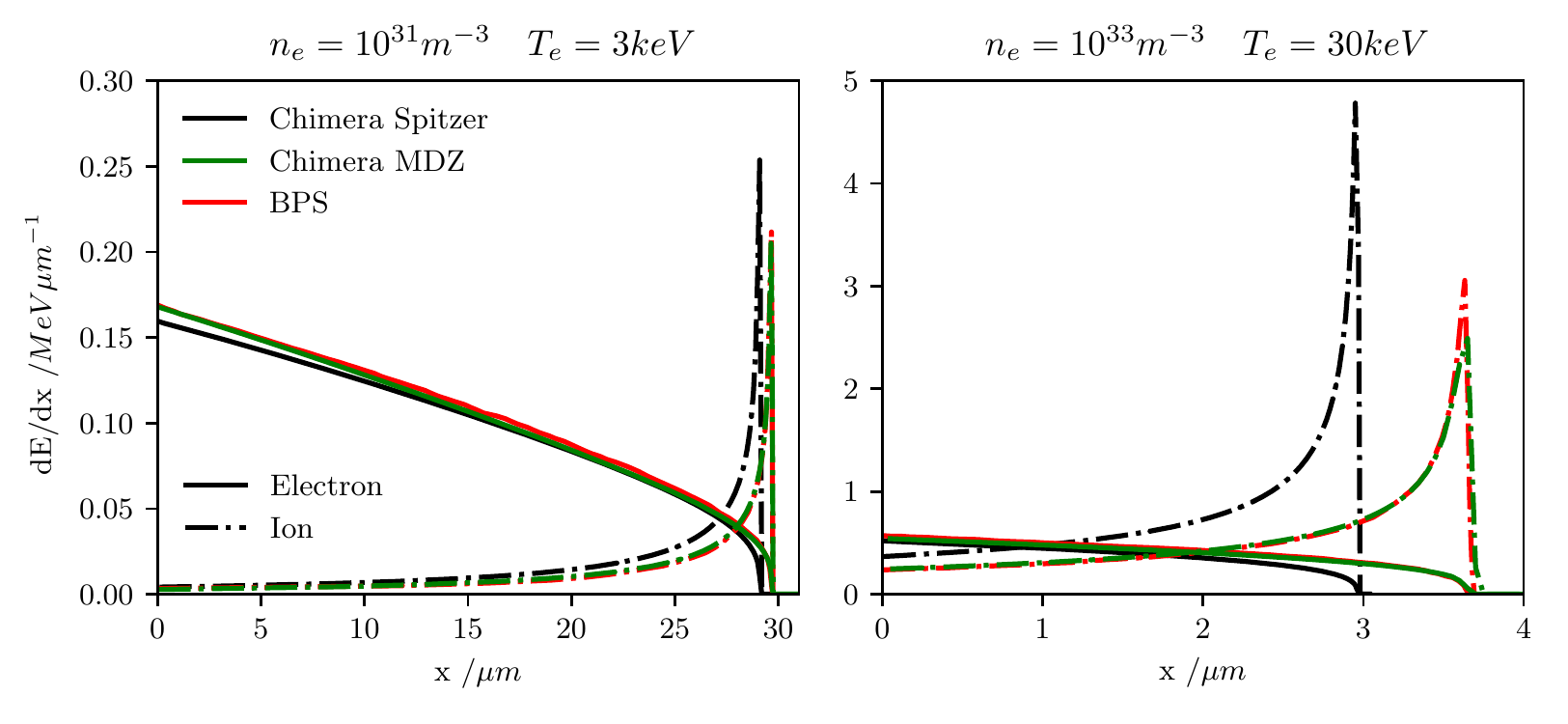}
	\caption{\label{fig:StoppingComp} A comparison of Spitzer (black) and MDZ (green) stopping powers from the Chimera burn module to the BPS (red) model in a uniform background equimolar DT plasma at (left) $n_{e} = 10^{31} m^{-3}$ (equivalent to ${\rho}_{DT} = 4.1 {\times} 10^4 kgm^{-3} $) and $T_{e} = 3keV$, and (right) $n_{e} = 10^{33} m^{-3}$ and $T_{e} = 30keV$, following Singleton \cite{Singleton2008}. The stopping power is separated into contributions from electrons (solid) and ions (dash-dot). The BPS model is found to agree well with experimental investigations \cite{Frenje2015,Cayzac2017}.}
\end{figure*}

\subsection{Stopping models}
Although an in-depth discussion of charged-particle stopping power is beyond the scope of this article, we  elaborate briefly on the stopping models currently implemented, beginning with the classical Spitzer treatment, in which the coefficients above (equations \ref{eq:SherlockSlow}-\ref{eq:SherlockDiffPerp}) are derived. In this approach, large-angle scattering and collective effects are ignored, and the Coulomb logarithm is cut-off using an impact parameter corresponding to a $ 90^{\circ} $ deflection \cite{Cohen1950}, $b_{min}\approx e^{2}/4\pi\epsilon_{0}m\left\langle v^{2}\right\rangle \approx e^{2}/(4\pi\epsilon_{0}\cdot3k_{b}T)$ and the Debye length, $b_{max}=\lambda_{e}=\sqrt{k_{b}T_{e}\epsilon_{0}/e^{2}n}$. Here, particles of species ${\alpha}$ travelling through a Maxwellian background of species ${\beta}$ experiences a energy loss rate of:
\begin{equation}
\frac{dE}{dx}=\frac{Z_{\alpha}^{2}Z_{\beta}^{2}e^{4}}{4\pi\epsilon_{0}^{2}}\frac{n_{\beta}\ln\Lambda_{\alpha\beta}}{m_{\alpha}v_{\alpha}^{2}}\left[\frac{m_{\alpha}}{m_{\beta}}\psi\left(y\right)-\psi'\left(y\right)\right]\label{eq:StopSpitzer}
\end{equation}
where $\psi\left(y\right)=2/\sqrt{\pi}\times\int_{0}^{y}\sqrt{\xi}e^{-\xi}d\xi$, $\psi'=d\psi/dy=2ye^{-y^{2}}/\sqrt{\pi}$, $y=v_{\alpha}^{2}/v_{\beta}^{2}$, $v_{\beta}^{2}=m_{\beta}/2kT_{\beta}$, and $\ln\Lambda_{\alpha\beta}$	is the Coulomb logarithm between the two species. This ($dE/dx$) is equal to $m \, dv_{{\alpha}{\parallel}}/dt$ from equation \ref{eq:SherlockSlow}.

The Maynard-Deutsch model \cite{Maynard1985}, which uses the random-phase approximation (RPA) to treat the dielectric function, provides a more advanced stopping model valid for arbitrary electron degeneracy. This model considers contributions from both free and bound electrons through Coulomb collisions and collective motion, and allows slowing calculations for any velocity ratio $v_{\alpha}/v_{e}$, but neglects ion contributions. We will proceed to use Zimmerman's parameterisation of the model \cite{Zimmerman1997} (hereafter labelled as MDZ), which is computationally tractable:	
\begin{align}
\frac{\partial E}{\partial x}&=\frac{Z_{\alpha}^{2}e^{4}}{4\pi\epsilon_{0}^{2}}\frac{1}{m_{e}v_{\alpha}^{2}}n_{F}L_{F}	\label{eq:StopMDZ} \\
L_{F}&=\frac{1}{2}\ln\left(1+\Lambda_{F}^{2}\right)\left(\mathrm{erf}\left(y\right)-\frac{2}{\sqrt{\pi}}ye^{-y^{2}}\right) \\
\Lambda_{F}&=\frac{4\pi m_{e}v_{e}^{2}}{h\omega_{pe}}\cdot\frac{0.321+0.259y^{2}+0.0707y^{4}+0.05y^{6}}{1+0.130y^{2}+0.05y^{4}} \\
v_{e} &=\begin{cases}
\sqrt{\pi}{h}({2\pi m_{e}})^{-1}\left[4n_{e}\left(1+e^{-\mu/T_{e}}\right)^{1/3}\right] & \textrm{degenerate \quad \, \,}\\
\sqrt{\left(2kT_{e}\right)/m_{e}} & \textrm{non-degenerate}
\end{cases}
\end{align}

We can compare the stopping power for both the Spitzer and MDZ models as implemented in Chimera with the Brown-Preston-Singleton (BPS) model \cite{Brown2005} in figure \ref{fig:StoppingComp}. The BPS model has been found to agree well with experimental investigations of ion stopping powers in plasmas at both high \cite{Frenje2015} and low velocity \cite{Cayzac2017}, but is computationally complex to calculate. Following Singleton \cite{Singleton2008}, we consider a 3.5MeV $\alpha$-particle travelling through a uniform background plasma of equimolar DT at (\textbf{a}) $T=3\text{keV}$ and $n_{e} = 10^{31} \text{m}^{-3}$ (equivalent to ${\rho}_{DT} = 4.1{\times} 10^{4} \text{kgm}^{-3}$ and (\textbf{b}) $T=30\text{keV}$ and $n_{e} = 10^{33} \text{m}^{-3}$. We have found our models (particularly MDZ) to be in good agreement in stopping power and range with BPS, particularly for conditions similar to those of an ICF hotspot figure \ref{fig:StoppingComp}a. As a default, we use the MDZ model owing to its excellent agreement with the BPS model, and ease of calculation. The diffusive coefficients are significantly smaller than the frictional slowing coefficient, and are calculated as equations \ref{eq:SherlockDiffPara} and \ref{eq:SherlockDiffPerp} in the Spitzer formulation, due to the lack of similarly computationally streamlined parameterisations of other formulations.

\subsection{Population Control}

As the production rate of alpha-particles changes rapidly in time, the computational weight (i.e. how many real particles are represented) of macro-particles being spawned is calculated dynamically.
As the fusion rate increases, the macro-particles being spawned will have increasingly higher weights, and correspondingly will be increasingly more important.
In addition, alpha-particles with thermal velocities contribute very little to heating, and thus thermalised particle motion does not need to be calculated.
In order to maintain computational and statistical accuracy, we need to create ${\sim}10^4-10^5$ macro-particles each time-step, with a total population limit (dictated by processor memory and simulation runtime) of ${\sim}10^6-10^7$.
With such a population limit where the simulation is likely to hit the cap, it is necessary to manage the macro-particle population in order to make sure we can continue to generate new particles where they are needed.

As such, we both terminate (i.e. remove from the simulation) any thermalised macro-particles, and also employ a basic population management scheme in our model.
Since the primary physical effect of note is each macro-particle's contribution towards the heating rate, which is affected by the macro-particle's energy and weight, we therefore expect the macro-particles with the highest total energy $E_{total}$ to have the most impact on the simulation; here, we define $E_{total}= w_{macro} \times E_{macro}$, where a macro-particle represents $w_{macro}$ particles of energy $E_{macro}$.
Hence we look to reduce the population by targetting those macro-particles with the lowest $E_{total}$.

We first dynamically calculate an energy threshold $E_{thresh}$ below which some fraction of particles, $1-f_{surv}$ will be removed from the simulation. We do this by using the maximum value of $E_{total}$, $E_{total, max}$ in the macro-particle population of size $N_{total}$, such that $E_{thresh}  = k E_{total, max}$ for some constant $k$. $k$ is increased from a default value of $0.1$ such that $f_{surv} {\in} (0,1)$, for:
\begin{equation}
f_{surv} = (N_{target} - N_{save}) / N_{PopControl}
\end{equation}
where $N_{target}$ is the target total population, $N_{save}$ is the number of macro-particles with $E_{total} > E_{thresh}$, $N_{PopControl}$  the number of macro-particles with $E_{total} \le E_{thresh}$, and $N_{save} + N_{PopControl} = N_{total}$. Once $k$ and $E_{thresh}$ are found, the population $N_{PopControl}$ is then reduced by a fraction $1 - f_{surv}$ through random annihilation of particles, with the $w_{macro}$ and $E_{total}$ of surviving $f_{surv}$ macro-particles increased by a factor of $1/f_{surv}$.

In effect, real particles and energy within the lowest total energy population are redistributed by removing macro-particles and increasing the weight and energy of surviving macro-particles  within this lowest total energy population in order to conserve the total energy and number of real particles.
Tests of the population control show no impact on the heating rates.

\subsection{Memory Structure}

As mentioned above, we need to create and terminate ${\sim}10^{4}-10^{5}$ macro-particles at every time-step with a total population of ${\sim}10^6-10^7$.
Thus, the primary demand on the data structure used to store our particle data is the ability to easily add and remove many elements (i.e. macro-particles, in this case) in any order, multiple times, and a secondary demand is being able to store a large number of particles. 
In addition, since the simulation domain is split across processors, we also need to be able to pass the information on the macro-particles between processors as they travel through the hotspot.

Owing to these requirements, we use a linked-list type data structure to store particle data rather than a dynamic array in order to optimise the memory-usage and computational speed of our model.
Linked lists are a collection of nodes containing `pointer' and `data' fields, with the pointer field containing the memory address of the next node.
In contrast, dynamic arrays are a collection of indexed elements, usually stored in a contiguous section of memory.

Linked lists allow for fast insertion and deletion of arbitrarily many elements (i.e. macro-particles), while array insertion and deletion requires either keeping track of every available index, or shifting all of the elements with every deletion and tracking the first available index.
Memory management is significantly easier, as the size of a linked list is limited only by the total memory available.
By comparison, dynamic arrays are limited by the length of contiguous memory available.
In addition, since the total number of macro-particles is not known \textit{a priori}, and indeed will change every time-step.
If the existing array becomes filled up, reallocating the data to a larger array is highly computationally expensive, particularly with increasingly large arrays. 
In addition, passing macro-particles between processors is easily achieved by moving elements to be passed to a new list, passing the entire list and then merging them.
Access to specific elements is difficult through linked lists, but easy for arrays due to the indexing. However, this feature is not needed, since we perform operations (such as pushing) on the entire list in a single-pass through.
In our model, linked lists allow for over ${\sim}100{\times}$ more macro-particles per processor than dynamic arrays, while also operating faster.

\newcommand{\newblock}{}
\bibliographystyle{unsrtnat}
\bibliography{./library_fixed,./Textbooks}

\end{document}